\documentclass{PoS}

\usepackage{amsmath,amssymb,amsbsy,mathbbol}

\title{GW Meson Scattering on a Staggered Sea}
\author{\speaker{Donal O'Connell}\thanks{I would like to thank my collaborators on this work, Jiunn-Wei Chen and Andr\'e Walker-Loud. \quad CALT 68-2611}\\
California Institute of Technology, 452-48, Pasadena, CA 91125, USA \\
Email: \email{donal@theory.caltech.edu}}

\abstract
{
We discuss the structure of the NLO corrections to the chiral formulae
for mesonic scattering processes in mixed action simulations using
Ginsparg-Wilson valence quarks and staggered sea quarks. In particular,
we show that the analytic contribution of the NLO chiral Lagrangian is
the same as in QCD. We also comment on how this result restricts the
dependence of the amplitudes on the unknown parameter $C_\mathrm{Mix}$
appearing in the chiral theory appropriate for these systems. We conclude
with some comments on the explicit scattering lengths.
}

\FullConference{XXIV International Symposium on Lattice Field Theory\\
July 23-28 2006\\
Tucson Arizona, USA}

\ShortTitle{GW Meson Scattering on a Staggered Sea}

\begin{document}

\section{Introduction}

In continuum quantum field theory in four dimensions, massless Dirac
fermions respect chiral symmetry. This symmetry is crucial for our
understanding of low energy processes of QCD: since the light quark
masses are small compared to the QCD scale, one would expect to see
signatures of chiral symmetry in the spectrum. Since these signatures are
not observed, we assume that QCD dynamically breaks chiral symmetry at
some scale $\Lambda_\chi$; this leads to concrete predictions which are
phenomenologically successful. Consequently, one would like latticized QCD
to enjoy some analogue of chiral symmetry. This is achieved if the quarks
satisfy the Ginsparg-Wilson relation~\cite{Ginsparg:1981bj}; current
realizations are domain wall fermions and overlap fermions. These quark
discretizations are collectively known as Ginsparg-Wilson (GW) quarks.

Unfortunately GW quarks come with a price. Simulations using GW quarks are
slower~\cite{Kennedy:2004ae} than simulations with Wilson or staggered
quarks. Staggered quarks, in particular, have been popular recently due
to the availability of the MILC lattices~\cite{Bernard:2001av}. But this
discretization scheme has its own issues: staggered chiral perturbation
theory~\cite{Lee:1999zx,Aubin:2003mg,Aubin:2003uc} is theoretically
complicated, as demonstrated, for example, by the large number of
operators in the NLO Lagrangian~\cite{ruth}.\footnote{We assume the
validity of the fourth-root trick, though it has not been shown that
the continuum limit of latticized QCD with rooted-staggered quarks is
in the same universality class as QCD.}

A practical compromise between the GW and staggered discretization
schemes is to use staggered fermions for the sea quarks and GW
fermions for the valence quarks\cite{Renner:2004ck}. This is
particularly attractive since one can employ the MILC lattices,
but as we shall see, it is also theoretically clean. Several recent
simulations~\cite{Beane:2005rj,Edwards:2005ym,Beane:2006pt,Beane:2006fk,Beane:2006kx,Beane:2006gj}
have utilized this kind of mixed action scheme, so it is important to
understand the chiral extrapolation appropriate to these simulations.

Nevertheless, these mixed action simulations face their own
challenges. Chiral perturbation theory for mixed action simulations
(MA$\chi$PT)~\cite{Bar:2002nr,Bar:2005tu} depends on a new, as yet
undetermined parameter, $C_\mathrm{Mix}$. This parameter shifts the
mass of mixed valence-sea mesons at tree level, so that for a meson with
valence quark $v$ and sea quark $s$,
\begin{equation}
m_{vs}^2 = B (m_v + m_s) + a^2 \Delta_{\mathrm{Mix}},
\end{equation}
where $\Delta_\mathrm{Mix} = 16 C_\mathrm{Mix} / f^2$ and $a$ is the
lattice spacing. In addition, a mixed action theory violates unitarity at
any finite lattice spacing, so one must take the continuum limit before
unitarity can be restored, unlike a partially quenched theory. We will
comment on these issues below.

Before we continue, let us introduce some notation. We shall denote
our valence quarks as $u, d$ and $s$ while our sea quarks are $j, l$
and $r$. The leading order (LO) mass of a meson of quark content $a,
b$ will be denoted $m_{ab}$. Meson masses which suffer a tree level
lattice correction will be denoted with a tilde; for example, the mass
of a meson with quark content $j,j$ is $\tilde m_{jj}^2 = 2 B m_j + a^2
\Delta_I$, where $a^2 \Delta_I$ is the mass shift of a taste identity
meson due to the staggered potential. The next-to-leading order (NLO)
pion mass will be denoted $m_\pi$ and similarly for the kaon mass. The
leading order decay constant is denoted by $f$ while the NLO pion and
kaon decay constants are $f_\pi$ and $f_K$, respectively.

\section{NLO Corrections}

Recently, the $\pi\pi$ scattering length was computed~\cite{us} in mixed action chiral perturbation theory (MA$\chi$PT) for GW valence quarks on a staggered sea. In terms of leading order parameters, the scattering length is given by
\begin{multline} \label{eq:pipiScattLenLO}
a_{I=2}^{(0)} = -\frac{m_{uu}}{8 \pi f^2} \Bigg\{ 1+\frac{m_{uu}^2}{(4\pi f)^2} \Bigg[
                      4 \ln \left( \frac{m_{uu}^2}{\mu^2} \right)               +4 \frac{\tilde{m}_{ju}^2}{m_{uu}^2} \ln \left( \frac{\tilde{m}_{ju}^2}{\mu^2} \right) 
             -1 +l^\prime_{\pi\pi}(\mu) \Bigg]      
    \\
- \frac{m_{uu}^2}{(4\pi f)^2} \Bigg[
\frac{\tilde\Delta_{PQ}^4}{6 m_{uu}^4} +\frac{\tilde\Delta_{PQ}^2}{m_{uu}^2} \left[ \ln \left( \frac{m_{uu}^2}{\mu^2} \right) +1 \right]
\Bigg]
+ \frac{\tilde\Delta_{PQ}^2}{(4\pi f)^2}l^\prime_{PQ}(\mu)    + \frac{a^2}{(4\pi f)^2} l^\prime_{a^2}(\mu)                \Bigg\},
\end{multline}
where $\Delta_{PQ}^2 = \tilde m_{jj}^2 - m_\pi^2$.
This result simplifies considerably~\cite{us} if one rewrites the scattering length in terms of the mass of the pion on the lattice and the decay constant of the pion measured on the lattice (we refer to these as lattice-physical parameters.) Then the scattering length becomes
\begin{equation} \label{eq:pipiScattLen}
        a_{I=2}^{(0)} = -\frac{m_\pi}{8 \pi f_\pi^2} \Bigg\{ 1                 + \frac{m_\pi^2}{(4\pi f_\pi)^2} \bigg[                        3\ln \left( \frac{m_\pi^2}{\mu^2} \right)                        -1 +l_{\pi\pi}(\mu) \bigg]
                -\frac{m_\pi^2}{(4\pi f_\pi)^2} \frac{\tilde\Delta_{PQ}^4}{6\, m_\pi^4}        \Bigg\}.
\end{equation}
It is worth commenting on some aspects of this simplification. In
Eq.~\eqref{eq:pipiScattLen}, the unknown counterterm which must be
determined on the lattice, $l_{\pi\pi}$, is identical to the continuum,
unquenched counterterm at NLO, while in Eq.~\eqref{eq:pipiScattLenLO}
there are additional unknown unphysical terms $l^\prime_{PQ}$ and
$l^\prime_{a^2}$. These unphysical effects must somehow be removed
in the analysis of the lattice data. The expression
Eq.~\eqref{eq:pipiScattLenLO} depends on the masses of the mixed
valence-sea meson, $\tilde m_{ju}$. This mass depends on the unknown parameter
$C_\mathrm{Mix}$; therefore, this parameter must be determined if
one wishes to extrapolate using Eq.~\eqref{eq:pipiScattLenLO}. The
lattice-physical expression Eq.~\eqref{eq:pipiScattLen}, however, does
not depend on this unknown parameter; in fact, the lattice-physical
expression does not depend in any way on the mixed valence-sea mesons,
and only differs from the continuum, unquenched scattering length by
a known term which has its origin in the unitarity violating flavour
neutral propagators. Our goal in this section is to explain why
these simplifications occur, and to what extent we may expect similar
simplifications in other mesonic processes.

To begin, let us consider the structure of the NLO Lagrangian of
MA$\chi$PT. This Lagrangian is determined by the symmetry structure of
the underlying partially quenched, mixed action form of QCD, supplemented
by the assumption that chiral symmetry is spontaneously broken. Since
the theory has two sectors --- the valence and sea sectors --- it is
worth considering the spurions arising from these sectors separately. In
the valence sector, the quarks satisfy the Ginsparg-Wilson relation;
therefore, they only explicitly violate chiral symmetry through the
quark mass term. This is the same as in continuum, unquenched QCD and
therefore the purely valence sector of the NLO chiral Lagrangian in
the mixed action theory is the Gasser-Leutwyler Lagrangian. There are
additional sources of chiral symmetry violation in the sea sector. In
particular, for staggered sea quarks, there are additional spurions
associated with taste symmetry violation at finite $a$. One might expect
that these spurions would lead to $a$ dependent analytic terms in the NLO
scattering amplitudes; however, we will now show that this does not occur.

At NLO in the power counting scheme appropriate to current simulations,
the NLO operators in the chiral Lagrangian only contribute at tree
level. Of course, all of the in and out states in a simulation involve
valence quarks, so for the purposes of an NLO computation one can set
the sea quarks equal to zero in the NLO operators. Since the spurions
arising from the sea sector go to zero when the sea quarks go to zero,
we see that at NLO the only relevant spurions are the purely valence
spurions. However, we have seen above that the NLO purely valence
Lagrangian coincides with the Gasser-Leutwyler Lagrangian, so no lattice
spacing dependent analytic terms can arise in this way. An exception
to this argument occurs for double-trace operators; however, these
operators can only renormalize the parameters $f$ and $B$ occurring in
the LO chiral Lagrangian. In lattice-physical parameters we eliminate $f$
and $B$ in favour of the decay constant measured on the lattice and the
masses of the particles on the lattice; this removes any lattice spacing
dependence arising from double trace operators. We conclude that there
can never be any lattice-spacing dependence arising from NLO operators in
mesonic scattering amplitudes, expressed in lattice-physical parameters,
at next-to-leading order.

Let us consider some simple examples to help make these arguments more
concrete. A list of the NLO operators appearing in the staggered chiral
Lagrangian is given in~\cite{ruth}. Each of these operators also appears
in the NLO Lagrangian describing mixed action chiral perturbation theory
with staggered sea quarks, with the rule that the taste matrices $\xi$
of Ref.~\cite{ruth} are only non-zero on the sea quark subspace; this
is equivalent to replacing $\Sigma \rightarrow P_s \Sigma P_s$, where
$P_s$ is the projector to the sea subspace, in spurions associated with
taste violation.  For example, one operator in the Lagrangian is
\begin{equation}
\mathcal{O}_1 = a^2 \mathrm{str} \left[ \partial_\mu \Sigma^\dagger \partial^\mu \Sigma \xi_5 P_s \Sigma^\dagger P_s \xi_5 P_s \Sigma P_s \right].
\end{equation}
We only want to use this operator at tree level, so we should take the sea
quarks to vanish; this sets $P_s \Sigma P_s \rightarrow 0$. Consequently,
$\mathcal{O}_1$ does not contribute at NLO. For another example, consider
the double trace operator
\begin{equation}
\mathcal{O}_2 = a^2 \mathrm{str} \left[ \partial_\mu \Sigma^\dagger \partial^\mu \Sigma \right] \mathrm{str} \left[\xi_5 P_s \Sigma^\dagger P_s \xi_5 P_s \Sigma P_s \right].
\end{equation}
In this case, it helps to remember that $\Sigma$ is given in terms of
the matrix of mesons $\Phi$ by $\Sigma = \exp [2 i \Phi / f]$, so when
we set the sea quarks to zero the trace over the sea subspace is just a
constant, namely the number of sea quarks $N_s$. Thus, for the purposes
of a next-to-leading order calculation $\mathcal{O}_2$ reduces to
\begin{equation}
\mathcal{O}_2 \rightarrow a^2 \mathrm{str} \left[ \partial_\mu \Sigma^\dagger \partial^\mu \Sigma \right] .
\end{equation}
Notice that the effect of $\mathcal{O}_2$ is to introduce an $a^2$
shift to the leading order parameter $f$.

A similar argument applies to potential counterterms involving sea
quark masses. These are associated with spurions arising from the sea
sector and therefore they may not contribute to scattering amplitudes
except to renomalise the parameters $f$ and $B$; this dependence
is removed upon switching to lattice-physical parameters. However,
there is another, perhaps more physical, way of seeing why the sea
quark masses do not contribute to the scattering. To understand this,
we must digress briefly on $\pi \pi$ scattering in $SU(3)$ chiral
perturbation theory. Evidently, the strange quark mass $m_s$ is a
parameter of $SU(3)$ $\chi$PT, and so one might think that the $\pi\pi$
scattering length includes analytic terms involving $m_s$. However,
if we suppose that the strange quark is heavy and integrate it out,
we must then recover $SU(2)$ $\chi$PT. Now, chiral symmetry forces any
$m_s$ dependence in the analytic terms of the $\pi \pi$ amplitude to
occur in the form $m_\pi^2 m_K^2$. But the only counterterm in
the on-shell $SU(2)$ scattering amplitude (Eq.~\eqref{eq:pipiScattLen}
with $\tilde \Delta_{PQ} = 0$) is proportional to $m_\pi^4$. It is not
possible to absorb $m_\pi^2 m_K^2$ into $m_\pi^4$, so there can be no
$m_s$ dependence in the $SU(3)$ $\pi\pi$ scattering amplitude. This is
indeed the case~\cite{Knecht:1995tr}.

Now, let us return to the situation with the valence-sea meson masses. For
the purposes of this discussion, we can ignore the flavour-neutral
and ghost sectors, reducing our theory from an $SU(6|3)$ theory to
an $SU(6)$ theory. The sea quark dependence of this $SU(6)$ chiral
perturbation theory is analogous to the $m_s$ dependence of $SU(3)$
$\chi$PT. A similar decoupling argument tells us that the sea quark
masses cannot affect processes involving the valence sector provided one
uses the analogues of on-shell parameters which are the lattice-physical
parameters. We conclude that there can be no analytic dependence on the
sea quark masses in a mesonic scattering amplitude. These rather abstract
arguments have been verified by explicit computation~\cite{forth}.

To conclude this section, we have shown so far that there can be no
dependence on the lattice spacing $a$ or the sea quark masses arising
from the NLO operators in the MA$\chi$PT Lagrangian at next-to-leading
order in the chiral power counting. However, the scattering amplitudes
can still depend on these quantities through loop corrections from the LO
Lagrangian. Note that since there is no counterterm available to absorb
divergences proportional to the lattice spacing or the sea quark masses;
this is a restriction on the form of the loop corrections. In favourable
cases this restriction is strong enough to remove all dependence on the
valence-sea mesons as occurred in the $\pi\pi$ amplitude; this removes
all dependence on the unknown parameter $C_\mathrm{Mix}$. In the next
section we will discuss individual mesonic processes and their structure.

\section{Meson meson scattering}

In a forthcoming work~\cite{forth} we present explicit expressions
for the $KK$ and $K\pi$ scattering amplitudes at threshold. Here, we
will restrict our comments to the dependence of these amplitudes on the
valence-sea mesons. First, let us introduce some notation. It is useful to
quantify the partial quenching in the mesonic sector with the parameters
\begin{align}
\Delta_{ju}^2 &= \tilde m_{jj}^2 - m_\pi^2 = 2 B (m_j - m_u) + a^2 \Delta_I \\
\Delta_{rs}^2 &= \tilde m_{rr}^2 - m_{ss}^2 = 2 B (m_r - m_s) + a^2 \Delta_I .
\end{align}
It is common to consider quantum theories which have unitary low energy
sectors but which violate unitarity above some cutoff $\Lambda$. For
example, $\phi^4$ theory with a Pauli-Villars regulator is simply a
theory or a light self-interacting scalar with a heavy ghost field which
violates unitarity. The parameters $\Delta$ which we have introduced are
measures of unitarity violation in the low-energy sector of QCD; when
these parameters are zero, for example, the effective theory in Minkowski
space has cuts at the points expected by the optical theorem. From
the point of view of the low energy theory, one would like to tune
these parameters to zero to remove as much of the unitarity violation as
possible. \footnote{At higher orders, the $\Delta$ parameters will receive
further corrections, but can generally be defined as the difference
between lattice-physical sea-sea meson masses and lattice-physical
valence-valence masses.} Note that this involves tuning the sea quarks
to be lighter than the valence quarks; the high energy theory of quarks
is non unitary but this is only seen above the non-perturbative scale
of QCD, when perturbative computations are possible. With these ideas
in hand, let us consider the mesonic scattering amplitudes.

The $KK$ scattering amplitude in lattice-physical parameters does not depend on the valence-sea mesons and hence the unknown parameter $C_\mathrm{Mix}$. When $\Delta_{rs} = 0$ the expression for the scattering amplitude is
\begin{align}
\mathcal{M} =&\ -\frac{4m_K^2}{f_K^2}
	+\frac{56 m_K^4}{9(4\pi)^2 f_K^4}
	-\frac{8 m_K^4}{(4\pi)^2 f_K^4}\, \ln \left( \frac{m_K^2}{\mu^2} \right)
	+\left( \frac{10 m_K^2 m_\pi^2}{9 (4\pi)^2 f_K^4} +\frac{2 m_\pi^4}{9 (4\pi)^2 f_K^4} \right)\, 
		\ln \left( \frac{m_\pi^2}{\mu^2} \right) \nonumber\\
	&\ -\left( \frac{m_\pi^4}{9 (4\pi)^2 f_K^4} +\frac{64 m_\pi^2 \tilde{m}_X^2}{45 (4\pi)^2 f_K^4}
		-\frac{46 m_K^2 \tilde{m}_X^2}{45 (4\pi)^2 f_K^4} +\frac{13 \tilde{m}_X^4}{5 (4\pi)^2 f_K^4}
		\right)\, \ln \left( \frac{\tilde{m}_X^2}{\mu^2} \right) \nonumber\\
	&\ -\frac{8(2m_K^2 +m_\pi^2)^2}{27(4\pi)^2 f_K^4 (\tilde{m}_X^2-m_\pi^2)}\,
		\left( \tilde{m}_X^2\, \ln \left( \frac{\tilde{m}_X^2}{\mu^2} \right)
			-m_\pi^2\, \ln \left( \frac{m_\pi^2}{\mu^2} \right) \right) + \frac{m_K^4}{f_K^4} \ell_{GL} (\mu) \nonumber\\
	&\ +\frac{\tilde{\Delta}_{ju}^2 \tilde{m}_X^2}{(4\pi)^2 f_K^4}\, \mathcal{F}_1\left( 
		\frac{m_\pi^2}{\tilde{m}_X^2}, \frac{m_K^2}{\tilde{m}_X^2}, \frac{\tilde{m}_X^2}{\mu^2} \right)
	+ \frac{\tilde{\Delta}_{ju}^4}{(4\pi)^2 f_K^4}\ \mathcal{F}_2\left( 
		\frac{m_\pi^2}{\tilde{m}_X^2}, \frac{m_K^2}{\tilde{m}_X^2}, \frac{\tilde{m}_X^2}{\mu^2} \right) 
	\nonumber\\
	&\ + \frac{\tilde{\Delta}_{ju}^6}{(4\pi)^2 f_K^4 \tilde{m}_X^2}\ \mathcal{F}_3 \left( 
		\frac{m_\pi^2}{\tilde{m}_X^2}, \frac{m_K^2}{\tilde{m}_X^2} \right)
	+\frac{\tilde{\Delta}_{ju}^8}{(4\pi)^2 f_K^4 \tilde{m}_X^4}\ \mathcal{F}_4 \left( 
		\frac{m_\pi^2}{\tilde{m}_X^2} \right), 
\end{align}
where the functions $\mathcal{F}_i$ are known functions~\cite{forth}. Note
that the counterterm $\ell_{GL}(\mu)$ in this expression is the same as the
counterterm in the physical scattering amplitude in on-shell parameters
as we expect from the discussion above.

The $K\pi$ scattering amplitude has a more complicated form than either
the $\pi\pi$ or $KK$ scattering amplitudes. 
This arises from the fact that the $\pi$ and $K$ meson masses are not
equal, which induces a net momentum flow through loop propagators in
the $u$-channel.
Consequently, in this case the scattering amplitude does
depend on the valence-sea mesons and therefore on $C_\mathrm{Mix}$. The
part of the $K\pi$ amplitude which depends on the sea quarks $F =
j,l,r$ is
\begin{equation}
\mathcal{M}_{vs} = \frac{1}{(4 \pi f^2)^2} m_K m_\pi \sum_F \left(
C_{Fd} \log \frac{\tilde m_{Fd}^2}{\mu^2} -
C_{Fs} \log \frac{\tilde m_{Fs}^2}{\mu^2} -
2 m_K m_\pi J(\tilde m_{Fd}^2) + 4 m_K m_\pi
\right)
\end{equation}
where
\begin{align}
C_{Fd} &= \frac{4 m_K m_\pi^2 - \tilde m_{Fd}^2 ( m_K + m_\pi)}
{m_K - m_\pi} \\
C_{Fs} &= \frac{4 m_K^2 m_\pi - \tilde m_{Fs}^2 ( m_K + m_\pi)}
{m_K - m_\pi}
\end{align}
and
\begin{equation}
J(m) = 4 \frac{\sqrt{m^2 - m_\pi^2}}{m_K - m_\pi} 
\arctan \left[
\frac{(m_K - m_\pi) \sqrt{m^2 - m_\pi^2}}{m^2 + m_K m_\pi -m_\pi^2}
\right] .
\end{equation}
Although the amplitude does depend on the valence sea-mesons, it is straightforward to check that the $\mu$ dependence of the amplitude does not depend on the these mesons, in agreement with our comments above. Of course, the counterterm in the full $K\pi$ amplitude in lattice-physical parameters is given by the QCD expression in on-shell parameters.

\section{Conclusions}

It is well-known that simulations with Ginsparg-Wilson valence quarks have
no lattice spacing dependence at tree level of chiral perturbation theory
as a consequence of the chiral properties of GW quarks. We have shown
that this persists at next-to-leading order in the chiral expansion for
mesonic scattering processes, expressed in lattice-physical parameters,
in the sense that the unknown analytic term which must be determined
from the lattice, does not depend on the lattice spacing. In addition,
we have shown that there is no unknown analytic terms involving the
sea meson masses. This situation is favourable because it simplifies
the extraction of the interesting physical information from a lattice
simulation of these processes: in principle, one need only use one
lattice spacing, for example.

In addition, we have commented on the dependence of the $KK$ and $K\pi$
scattering amplitudes on the unknown parameter $C_\mathrm{Mix}$ which
appears in the leading order chiral Lagrangian describing mixed action
processes. The $KK$ scattering amplitude does not depend on this
parameter, but the $K\pi$ amplitude does; therefore, correct chiral
extrapolation of $K\pi$ scattering data requires the determination of
this parameter.


\end{document}